\documentclass[longnamesfirst,apj]{emulateapj}
\usepackage{apjfonts}

\newcommand{\beq}{\begin{equation}}
\newcommand{\eeq}{\end{equation}}
\newcommand{\bea}{\begin{eqnarray}}
\newcommand{\eea}{\end{eqnarray}}

\begin{document}
\title{An Eccentric Circumbinary Accretion Disk and the Detection of  Binary Massive Black Holes}
\author{Andrew I.~MacFadyen\altaffilmark{1,2} and Milo\v s Milosavljevi\'c\altaffilmark{3,4,5}}
\altaffiltext{1}{Institute for Advanced Study, Einstein Drive, Princeton, NJ 08540.}
\altaffiltext{2}{Department of Physics, New York University, New York, NY 10003.}
\altaffiltext{3}{Theoretical Astrophysics, Mail Code 130-33, California Institute of Technology, 1200 East California Boulevard, Pasadena, CA 91125.}
\altaffiltext{4}{Hubble Fellow.}
\altaffiltext{5}{Department of Astronomy, University of Texas, 1 University Station C1400, Austin TX 78712.}
\righthead{CIRCUMBINARY MASSIVE BLACK HOLE ACCRETION DISK}
\lefthead{MACFADYEN \& MILOSAVLJEVI\'C}
\begin{abstract}

We present a two-dimensional grid-based hydrodynamic simulation of a thin,
viscous, locally-isothermal corotating disk orbiting an equal-mass Newtonian
binary point mass on a fixed circular orbit.  We study the structure of the
disk after multiple viscous times.  The binary maintains a central hole in
the viscously-relaxed disk with radius equal to about twice the binary
semimajor axis. Disk surface density within the hole is reduced by orders of
magnitude relative to the density in the disk bulk. The inner truncation of
the disk resembles the clearing of a gap in a protoplanetary disk.  An
initially circular disk becomes elliptical and then eccentric.  Disturbances
in the disk contain a component that is stationary in the rotating frame in
which the binary is at rest; this component is a two-armed spiral density
wave.  We measure the distribution of the binary torque in the disk and find
that the strongest positive torque is exerted inside the central low-density
hole.  We make connection with the linear theory of disk forcing 
at outer Lindblad resonances (OLRs) and find that the measured torque density
distribution is consistent with forcing at the 3:2 ($m=2$) OLR, 
well within the central hole.  
We also measure the time dependence of the
rate at which gas accretes across the hole and find quasi-periodic
structure. We discuss implications for variability and detection of active
galactic nuclei containing a binary massive black hole.


\end{abstract}

\section{Introduction}
\label{sec:intro}

Circumbinary disks have been observed in young stellar binaries, such as the
spatially-resolved pre--main-sequence binary star GG Tau (e.g.,
\citealt{McCabe:02,Krist:05}).  Circumbinary disks should also form around
binary massive black holes when interstellar gas accretes into the binary's
dynamical sphere of influence (e.g.,
\citealt{Artymowicz:96,Ivanov:99,Armitage:02,Milosavljevic:05}).  It is
generally expected that the dynamics of a binary embedded in an accretion
disk is similar to the dynamics of a planet embedded in a protoplanetary
disk.  The planet orbits the central star while exchanging gravitational
torques with, and accreting from, a gaseous circumstellar disk.  If the
radius of the planet's Roche lobe exceeds the pressure scale height of the
disk, the planet clears a gap in the disk
(e.g.,~\citealt{Goldreich:80,Takeuchi:96,Lubow:99}).  When a similar-mass
binary is surrounded by a thin corotating disk, this gap-opening condition is
automatically satisfied.

An understanding of circumbinary accretion flows is essential for future
identification of binary massive black holes in astronomical surveys. Massive
binary black holes have only tentatively been identified in astronomical
observations (\citealt{Komossa:03,Komossa:06,Rodriguez:06} report evidence
for a binary with projected separation of $7.3\textrm{ pc}$), but the
theoretical case for their existence is very strong.  A massive binary black
hole forms during the conclusion of a galaxy merger and persists at parsec or
subparsec separations over a cosmological time. Under propitious
circumstances, gravitational radiation ultimately induces coalescence between
the black holes \citep[and references therein]{Begelman:80,Merritt:05}.

At parsec separations, the black holes can be optically resolved only in
nearby galaxies. Better resolution can be achieved at radio frequencies, but
binary radio cores can be confused with structures in radio jets and chance
superpositions. An unresolved binary can be detected by associated
characteristic spectral and time-dependent signatures if a circumbinary
accretion flow is present.  For example, asymmetries in the accretion flow
and time-dependent accretion onto the component black holes are sources of
variability that can be detected in monitoring surveys such as with the 
Large Synoptic Survey Telescope (LSST)\footnote{http://www.lsst.org}.

To predict the spectral signatures of massive binary black holes, one must
analyze the hydrodynamic structure of the circumbinary accretion flow and
then synthesize the radiative spectra emitted by the flow.  In this work we
carry out the hydrodynamic analysis in the case of a thin corotating disk.
We defer the spectral synthesis to subsequent work (see also
\citealt{Milosavljevic:05,Bogdanovic:07b}).  The search for massive binary
black hole spectral transients is most effective if one can identify robust
transients that repeat continuously over an extended period.  Transients that
are restricted to a single moment in the life of a binary, such as its
formation, will be rare and difficult to interpret.  Quasi--steady-state
circumbinary accretion disks are particularly attractive as they may persist
long enough to render their astronomical detection probable.

Previous numerical simulations of circumbinary disks based on the smoothed particle hydrodynamics method
\citep{Artymowicz:94,Escala:05} and grid-based hydrodynamics with explicit
bulk viscosity \citep{Guenther:02} suggest that the surface density of a
circumbinary disk drops sharply inward of $r\sim 1.5-2a$, where $a$ is the
binary semimajor axis.  The binary resides within a low-density hole,
also called the circumbinary gap, at the center of the disk.  Although
the gas density within the hole is low, some mass transfer does take place
from the disk onto the point masses \citep{Guenther:04}.  However the
accretion rate is below the rate that would be expected neglecting the torque
that the binary exerts on the disk.  The reduced accretion was detected in
simulations of massive planets embedded in protoplanetary disks (e.g,
\citealt{Lubow:99,Lubow:06}).

The simulations of disks with planets also indicate that an initially
circular disk becomes eccentric over time, even if the orbit of the planet is
circular \citep{Papaloizou:01,Kley:06}.  The growth of disk eccentricity has
been attributed to a dynamical instability \citep{Lubow:91,Dangelo:06} driven
by the planet's tidal potential.  In turn, the disk eccentricity can excite
eccentricity in the binary, since the two eccentricities are coupled
\citep{Papaloizou:01}.  Even a weak residual eccentricity can be detected in
the gravitational radiation emitted during black hole coalescence
(\citealt{Armitage:05}; see also \citealt{Artymowicz:91}). Therefore an
understanding of the circumbinary disk dynamics is key for electromagnetic
and gravitational wave observations of coalescing massive black holes with
the {\it Laser Interferometer Space Antenna}
({\it LISA}).\footnote{http://lisa.nasa.gov} Simultaneous electromagnetic and
gravitational detection can be used to measure the geometry of the universe
and constrain properties of dark energy (e.g., \citealt{Holz:05,Kocsis:06}).

We present high resolution simulations of a corotating viscous circumbinary
disk around a circular Newtonian equal-mass binary.  The disk lies in the
orbital plane of the binary. This is an astrophysically-motivated simplifying
assumption; alignment of the orbital plane of the inner disk and that of the
binary will take place because differential precession of a non-aligned disk
caused by the binary's mass quadrupole moment forces the inner part of the
disk to align with the orbital plane of the binary
(e.g.,~\citealt{Larwood:97}).\footnote{The orbital plane of the binary will
itself precess due to the relativistic spin-orbit coupling, if the binary
components have spins misaligned with the binary's angular momentum.}

In \S~\ref{sec:simulations} we describe the computational method and the
binary-disk setup.  In \S~\ref{sec:results} we present our results.  We
study the disk eccentricity, the spatial distribution of torque deposition in
the disk, and the mass accretion across the central hole in the disk.  In
\S~\ref{sec:discussion} we compare the torque 
distribution measured in the simulation 
with predictions of the theory of linear response of tidally forced disks.  
We also present a tentative interpretation of the eccentricity excitation
in terms of a mean motion resonance.  In \S~\ref{sec:black_holes} we discuss
implications for astronomical detection of binary massive black hole mergers
in progress.  Finally, in \S~\ref{sec:conclusions} we summarize the main
conclusions.

\section{The Code and Initial Conditions}
\label{sec:simulations}

The simulation was carried out with the publicly available code ASC FLASH
\citep{Fryxell:00} in two spatial dimensions. The mesh refinement
capability of FLASH allowed us to use high grid resolution in the central
region of the disk (within a radius equal to 10 binary semimajor axes) and
lower resolution in the outer disk.  We implemented a static grid with
constant resolution in the $\phi$ direction.  The piecewise-parabolic method
module implemented in FLASH is ideally suited for treating supersonic
dynamics in the vicinity of the binary.  The simulation was carried out in
polar ($r$,$\phi$) coordinates in the inertial frame.

We implemented several small changes in FLASH. We introduced a Newtonian
binary point particle of total mass $M$ on a fixed circular orbit centered at
the origin.  The gravity of the binary acting on the disk fluid is evaluated
directly; the gravity of the fluid is not calculated. The integration domain
is a circular annulus with an inner boundary at $r_{\rm min}=a$ and outer
boundary at $r_{\rm max}=100a$.  The binary point masses orbit at the
distance $a/2$ from the center; therefore the point masses move entirely
within the excised circular region $r<r_{\rm min}$.  An obvious limitation
implied by the excision is that we do not study the accretion of gas onto the
individual point masses, but we do measure the net rate of mass transfer
across the inner boundary.  

The fluid was allowed to flow freely out of the
integration domain at the inner and outer boundary, but the flow into the
domain was forbidden.  Specifically, the boundary conditions at  
$r_{\rm min}$ and $r_{\rm max}$ were such that the radial derivative of the surface density, $\partial\Sigma/\partial r$, and of the azimuthal velocity, $\partial v_\phi/\partial r$, were set to vanish at each boundary. The boundary condition imposed on the radial velocity $v_r$ was a function of the direction of the flow at the boundary. For outflow, $v_r(r_{\rm min})<0$ or $v_r(r_{\rm max})>0$, the radial derivative of the radial velocity, $\partial v_r/\partial r$, was set to vanish; for inflow, the value of $v_r$ in the boundary (``ghost'') zones just adjacent to $r_{\rm min}$ and $r_{\rm max}$ was set to vanish, thus preventing inflow.

We implemented a locally isothermal equation of state in which the sound
speed is a fixed function of radius $c_{\rm s}(r)$.  The disk has a constant
pressure scale height to radius ratio $h/r=c_{\rm s}/r\Omega_{\rm kep}=0.1$,
where $\Omega_{\rm kep}=(GM/r^3)^{1/2}$ is the Keplerian frequency in the
disk; therefore the sound speed decreases with radius $c_{\rm s}\propto
r^{-1/2}$. Since this equation of state is applied exactly, the total
energy of the fluid is not conserved and all shocks can be considered
radiatively isothermal.

We also implemented an explicit viscous shear stress term in the equation
governing the transport of the $\phi$-momentum.  In doing so, we kept only
the $(r,\phi)$ component of the stress tensor, since this is the component
that communicates viscous torque in the disk and is responsible for
accretion.  The corresponding vertically-integrated Navier-Stokes equation
reads \bea
\label{eq:ns_azimuthal}
\frac{\partial v_\phi}{\partial t}+
\frac{v_r}{r}\frac{\partial (rv_\phi)}{\partial r}&+&
\frac{v_\phi}{r}\frac{\partial v_\phi}{\partial\phi}-\frac{\partial \Phi}{r\partial \phi}
\nonumber\\
&+&
\frac{1}{\Sigma}\left[\frac{\partial P}{r\partial\phi}
+ \frac{\partial(r^2\sigma_{r\phi})}{r^2\partial r}\right]=0,
\eea
where $\Sigma$ is the surface density in the disk,
\bea
\label{eq:potential}
\Phi(r,\theta) &=&-\frac{\onehalf GM}{[r^2+\onequarter a^2-ra\cos(\theta-\Omega_{\rm bin}t)]^{1/2}}
\nonumber\\& &
- \frac{\onehalf GM}{[r^2+\onequarter a^2+ra\cos(\theta-\Omega_{\rm bin}t)]^{1/2}}
\eea
is the Newtonian gravitational potential of the binary point mass,
$\Omega_{\rm bin}\equiv(GM/a^3)^{1/2}$ is its angular frequency, $P=\Sigma
c_{\rm s}^2$ is the vertically-integrated pressure, and $v_{r}$ and
$v_{\phi}$ are the radial and the tangential component of the fluid velocity,
respectively.

The $(r,\phi)$ component of the stress tensor equals
\beq
\label{eq:stress_general}
\sigma_{r\phi}=\Sigma\, \nu\left[r\, \frac{\partial}{\partial r}
\left(\frac{v_\phi}{r}\right)
+\frac{1}{r}\frac{\partial v_r}{\partial\phi}\right] ,
\eeq
where $\nu$ is the kinematic viscosity.  On the right-hand side of equation
(\ref{eq:stress_general}), the second term involving $\partial
v_r/\partial\phi$ is normally much smaller than the first term involving
$\partial(v_\phi/r)/\partial\phi$. We drop the second term in the simulation.

The viscosity is calculated based on the $\alpha$-disk prescription
$\nu(r)=\alpha c_{\rm s}(r)^2/\Omega_{\rm kep}(r)$ with $\alpha=0.01$. The
use of the Keplerian frequency in the evaluation of the viscosity is simply a
prescription; the binary's mass quadrupole moment and positive radial
pressure gradients make the true angular frequency near the inner edge of the
disk super-Keplerian
\beq
\label{eq:omega}
\Omega^2\approx \Omega_{\rm kep}^2 \left(1+\frac{3}{16}\frac{a^2}{r^2}\right)^2+
\frac{1}{r\Sigma} \frac{dP}{dr} .
\eeq

Numerical resolution in the simulation
was chosen to ensure an accurate treatment of pressure
waves excited in the disk.
The binary's tidal potential can excite density waves at circular and
eccentric outer Lindblad resonances (OLRs).  
These waves have wavelengths equal to a
rational multiple of $\sim 2\pi c_{\rm s}(r)/\Omega_{\rm bin}\sim 0.6\ a
(r/a)^{-1/2}$.  The radial resolution 
was chosen such that the wavelength of a density wave is resolved by
many resolution elements 
in the inner disk (within a few binary separations from the inner
edge).  Specifically, the simulation was carried on time-independent (non-adaptive) nonuniform meshes utilizing the PARAMESH adaptive mesh refinement package in FLASH. The spatial resolution was ($\Delta r \approx 0.039\ a$, $\Delta\phi\approx 0.0078\times2\pi$) near the inner boundary ($r\sim r_{\rm min}$) and eight times lower ($\Delta r\approx 0.31\ a$, $\Delta\phi\approx 0.062\times2\pi$) near the outer boundary ($r\sim r_{\rm max}$). The resolution was uniform in the azimuthal direction at any fixed radius.

\begin{figure}
\epsscale{1.1}
\plotone{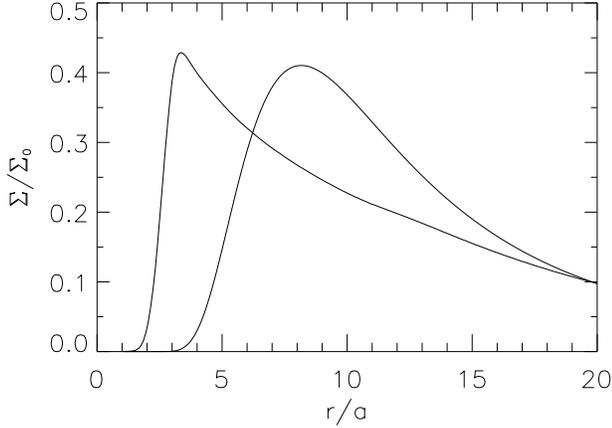}
\caption{Surface density of the disk a function of radius in units of the
binary semimajor axis. The curve peaking on the right is the initial disk
density profile; the one peaking on the left is density profile after $\sim
4,000$ binary orbits averaged over $\sim 50$ orbits.\label{fig:density}}
\end{figure}

The initial disk has radial surface density profile 
\beq
\label{eq:initial_disk}
\Sigma(r)=\Sigma_0 \left(\frac{r}{r_{\rm s}}\right)^{-\delta}\exp\left[-\left(\frac{r}{r_{\rm s}}\right)^{-\xi}\right] ,
\eeq
where $r_{\rm s}=10a$, $\delta=3$, $\xi=2$, and $\Sigma_0$ is an arbitrary
constant.  The initial density profile is shown in Figure
\ref{fig:density}. The surface density decreases to a small value near the
boundaries of the integration domain at $r_{\rm min}=a$ and $r_{\rm
max}=100\,a$.

The initial angular velocity of the fluid is corrected for the radial pressure gradient (see equation \ref{eq:omega})
and the initial radial velocity is corrected for the viscous drift
$v_r\approx 2(r^2\Omega \Sigma)^{-1} \partial(r^2\sigma_{r\phi})/\partial r$.
The initial velocity of the disk correctly represents the quasi-steady state of an axisymmetric disk to the first order in $h/r$ and $\alpha$, and no significant transients are observed at the beginning of the simulation. 

Since the inviscid hydrodynamical component of FLASH has been extensively
tested by other authors \citep{Fryxell:00}, we are here mainly concerned with
the correctness and stability of our implementation of explicit viscosity.
To test the implementation, we simulated an identical initial accretion disk
surrounding a single point particle of mass $M$ at the origin. This disk was
simulated at the same resolution as the circumbinary disk.  The test disk
remains axisymmetric and its evolution follows that prescribed by equation
(\ref{eq:ns_azimuthal}) on a viscous timescale, thereby validating our
implementation of the explicit viscosity.  

Maintaining exact axisymmetry for many thousands of orbits, as required for
simulations lasting many viscous times, required care with the numerical
grid.  We set the FLASH parameter {\tt unbiased\_geometry}
to clean the last bits
of floating point numbers defining the numerical mesh.  This was necessary 
because by default, due to numerical roundoff and truncation errors, 
the mesh spacing in the azimuthal direction is not precisely uniform. This
nonuniformity can seed nonaxisymmetric 
perturbations on many orbital timescales; 
{\tt unbiased\_geometry} enforces exact uniformity in the 
azimuthal mesh spacing, thus
eliminating spurious loss of
axisymmetry which developed in test simulations run for a single point mass
with and without viscosity.

\section{Results}
\label{sec:results}

\subsection{Surface Density Profile}

The initial disk given in equation (\ref{eq:initial_disk}) is smoothly
truncated at $r\sim 3a$.  The binary tide experienced by the initial disk
is weak and the disk initially accretes inward on a viscous time scale \beq
t_{\rm visc}\sim \frac{r^2}{3\nu}\sim
\frac{1}{3}\left[\alpha\left(\frac{h}{r}\right)^2\Omega\right]^{-1}\sim 1,600
\left(\frac{r}{2a}\right)^{3/2} t_{\rm bin} , \eeq where $t_{\rm bin}\equiv
2\pi/\Omega_{\rm bin}$ is the orbital period of the binary. Near the inner
edge the disk responds significantly faster to the viscous torque because of
the presence of a steep density gradient.  As the disk accretes inward toward
the binary, the binary's tidal forcing of the inner edge of the disk
strengthens.  When the viscously spreading disk arrives at a radius $r\sim
2a$, strong nonlinear disturbances develop in the circumbinary flow; we
discuss these disturbances in detail below.

\begin{figure}
\plotone{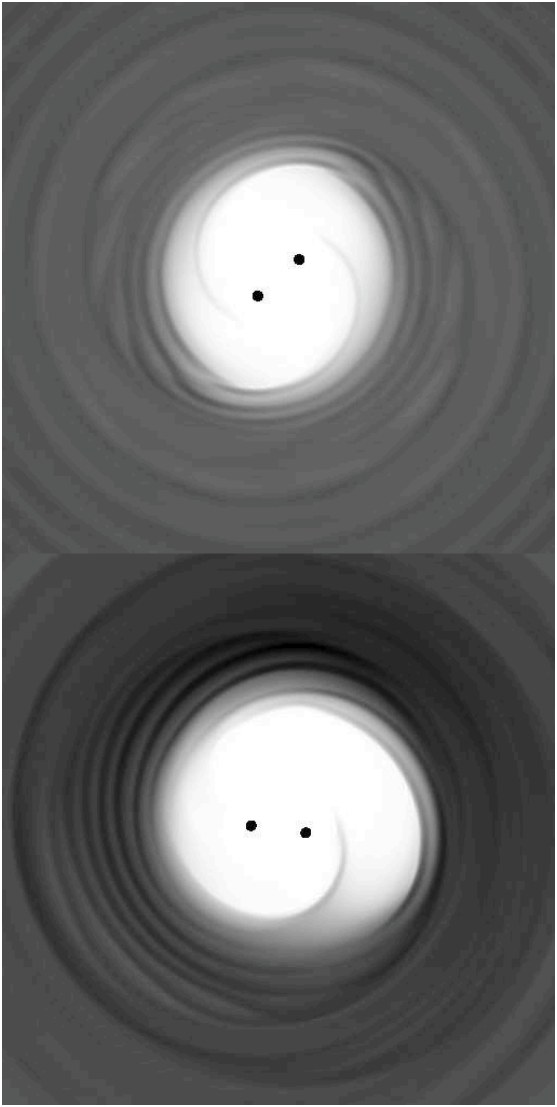}
\caption{Disk surface density at $1000$ binary revolutions ({\it top panel})
 and at $4000$ revolutions ({\it bottom panel}); the black dots indicate the
 location of the point masses.
 The plotted region spans $x/a=\pm 5$ and $y/a=\pm 5$, showing the inner
 $20\%$ (in radius) of the simulation region. \label{fig:2d}}
\end{figure}

The disk then settles into a quasi-steady state.  Figure \ref{fig:density}
shows the initial surface density profile and the time-averaged surface
density profile after 4000 orbits.\footnote{Here and elsewhere in the text
time averaging is carried out over $\sim50$ binary orbits.} The evolved disk
is sharply truncated inward of $r\sim 2a$ and the central region $r<2a$
contains very little fluid. The surface density in this region is an exponential
function of radius,
\beq
\label{eq:sigmasmall}
\langle\Sigma\rangle(r<2a) \approx \Sigma_0 \exp(-15.2+5.95\ r/a) ,
\eeq  
where 
\beq
\langle\cdot\rangle \equiv\frac{1}{2\pi} \int_0^{2\pi} d\phi
\eeq 
denotes the azimuthal average.  While the azimuthally averaged inner density
is low, accretion still occurs in streams of gas (Fig.~\ref{fig:2d})
which have radially infalling velocity components.

\subsection{Eccentricity and Ellipticity of the Disk}

\begin{figure}
\epsscale{1.1}
\plotone{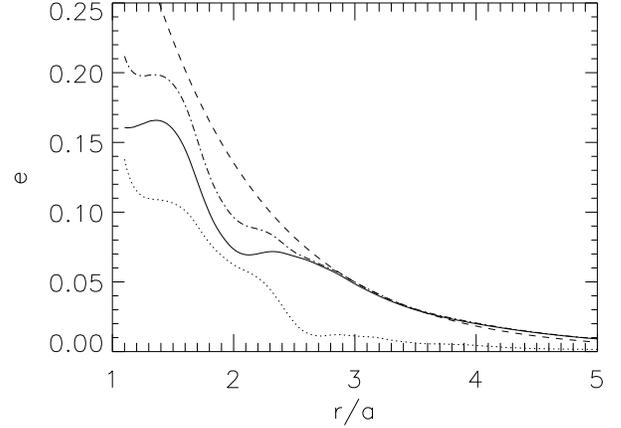}
\caption{Time-averaged disk eccentricity $\tilde e_1$ ({\it solid line}) and ellipticity $\tilde e_2$ ({\it dotted line}) as a function of radius after 4,000 binary orbits.  The two added in quadrature $(\tilde e_1^2+\tilde e_2^2)^{1/2}$ are also plotted ({\it dot-dashed line}).  Dashed thin line is the function $\exp(-r/a)$. \label{fig:eccentricity}}
\end{figure}

Dynamics of the inner disk is strongly nonlinear and time-dependent.  The
disk first becomes elliptical (Fig.~\ref{fig:2d}, top) and then
eccentric (Fig.~\ref{fig:2d}, bottom).  The elliptical state is sustained only on a single
viscous timescale, but the subsequent eccentric state is persistent and
apparently represents a quasi-steady state of the circumbinary disk.
Therefore, we focus on the eccentric disk in the following discussion.
We measure the magnitude of the
eccentricity, denoted by $\tilde e_1$, and the disk ellipticity, denoted by
$\tilde e_2$, by taking a time average of the density-weighted $m=1$ and
$m=2$ components, respectively, of the radial velocity, normalized by the
density-weighted azimuthal velocity 
\beq 
\tilde e_m\sim \frac{\left| \langle
\Sigma v_r e^{im\phi}\rangle \right|}{\langle \Sigma v_\phi\rangle} .  
\eeq
Figure \ref{fig:eccentricity}, shows the final, quasi--steady-state disk
eccentricity and ellipticity as a function of radius.  For $1.5a\lesssim
r\lesssim 5a$, the quantity $(\tilde e_1^2+\tilde e_2^2)^{1/2}$ is
roughly approximated by the function $\exp(-r/a)$.  The eccentricity of the
fluid in the central hole is nonlinear in the sense that the associated
epicyclic motion is (mildly) supersonic.

The disk
eccentricity precesses slowly in the prograde direction, and the precession appears to be rigid at radii $2a<r<4a$ . 
The average precession rate of the eccentricity vector (the line of apsides) at these radii is\footnote{In addition to the linear trend $\sim \Omega_{\rm prec}t$, the measured line of apsides executes quasi-periodic oscillations.}
\bea
\Omega_{\rm prec} &\equiv& \frac{d}{dt}\ {\rm arg}\ \langle\Sigma v_r e^{i\phi}\rangle\nonumber\\&\approx& 3.05\times10^{-3}\ \Omega_{\rm bin} .
\eea
This can be compared with the apsidal precession rate $\dot\varpi$ that would be expected for a test particle orbiting in the gravitational potential of the binary
\beq
\label{eq:pomega}
\dot\varpi = - \frac{1}{2r^2\Omega_{\rm kep}}
\frac{d}{dr} \left(r^2 \frac{d\tilde \Phi}{dr}\right) ,
\eeq
where
\beq
\label{eq:nonkeplerian}
\tilde \Phi\equiv \langle \Phi \rangle + r^2 \Omega_{\rm kep}^2
\eeq
is the phase-averaged non-Keplerian contribution to the gravitational potential.  Inserting the gravitational potential of the binary (equation \ref{eq:potential}) in equations (\ref{eq:nonkeplerian}) and (\ref{eq:pomega}) yields
\bea
\dot\varpi &=& \frac{3}{16} \left(\frac{r}{a}\right)^{-2}  \Omega_{\rm kep} +  {\cal O} \left(\frac{r}{a}\right)^{-4} \Omega_{\rm kep} \nonumber\\
&\approx& 4\times10^{-3}\ \left(\frac{r}{3a}\right)^{-7/2} \Omega_{\rm bin} .
\eea
The measured, radially-averaged precession rate $\Omega_{\rm prec}$ agrees with the test-particle apsidal precession rate $\dot\varpi$ at radius $r=3.24\ a$, which falls within the range of radii where the disk appears to precess as a rigid body, and is close to the radius where the disk surface density peaks and most of the mass of the rigidly-precessing disk resides (cf.\ Figure \ref{fig:density}). Therefore, the precession of the inner disk can be understood as arising from the non-Keplerian nature of the binary's gravitational potential. Exact agreement between $\Omega_{\rm prec}$ and $\dot\varpi$ may not be expected at any specific radius because of the role of radial pressure gradients, which are neglected in the test-particle approximation, that stabilize the rigid eccentric distortion.

\subsection{Torque Density}

\begin{figure}
\epsscale{1.1}
\plotone{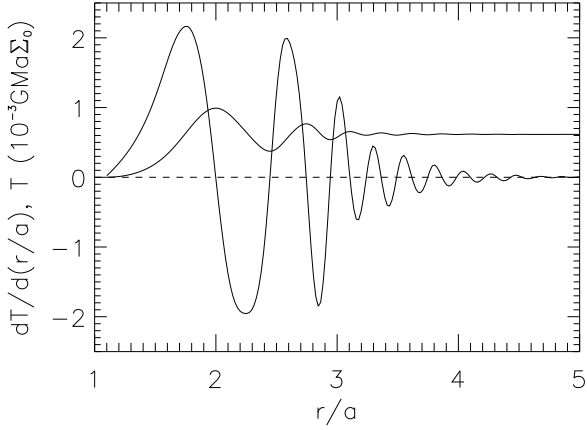}
\caption{Time-averaged torque density $dT/dr$ and the integrated torque $T$ exerted by the binary on the disk at $\sim4000$ binary revolutions.  Time averaging was carried out over $45$ binary revolutions, and reflects a quasi-steady state.   The total torque experienced by the disk, $T(\infty)$, is positive.\label{fig:torque}}
\end{figure}

The binary torques the disk by coupling gravitationally to nonaxisymmetric
structures in the disk.  In Figure \ref{fig:torque}, we plot the
time-averaged torque density
\beq
\frac{dT}{dr}\equiv - 2\pi r\left\langle\Sigma \frac{\partial\Phi}{\partial\phi}  \right\rangle,
\eeq
 and the integrated torque
\beq  
T \equiv \int_a^{r} \frac{dT}{dr} dr ,
\eeq 
where, as before, $\Phi$ denotes the gravitational potential of the binary.
The torque density can be of either sign. The strongest positive torque density is exerted at $r < 2 a$.  In this region the fluid density is relatively low, but the tidal force accelerates the fluid radially to highly supersonic velocities $v_r\sim v_\theta\gg c_{\rm s}$.  A weaker, {\it negative} torque density is exerted at $2a<r<2.4a$; at larger radii the torque density oscillates around the zero point.

\begin{figure}
\epsscale{1.1}
\plotone{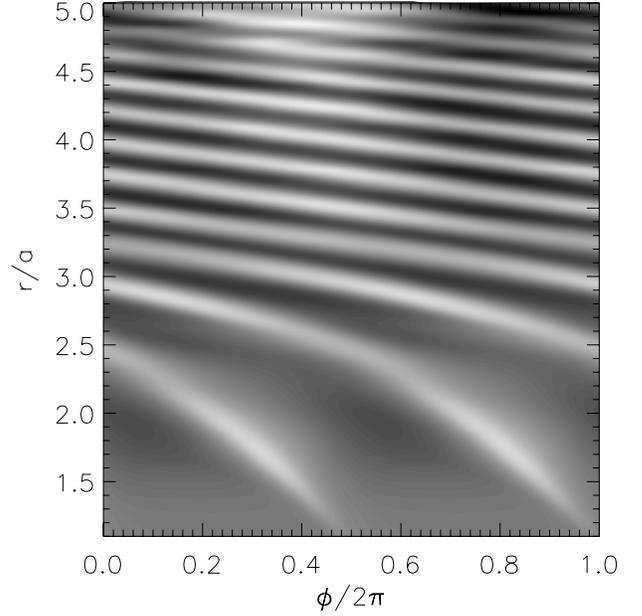}
\caption{Time-averaged surface density fluctuation $\delta\Sigma\equiv
  (\Sigma-\langle\Sigma\rangle)/\langle\Sigma\rangle$ in the rotating frame
  in the which the binary is at rest.  We have multiplied $\delta\Sigma$ by
  $(r/a)^7$ for clarity. The binary point masses are located at $r=\onehalf a$ and $\phi=(0,\pi)$.  The brightest regions are density maxima. The plot does not show the disk at radii $r_{\rm min}<r<1.1a$.\label{fig:wave}}
\end{figure}

To identify the origin of the oscillations in the torque density, we work in the rotating frame in which the binary is at rest and carry out time averaging of the surface density in this frame.  Figure \ref{fig:wave} shows the surface density fluctuation
\beq
\delta\Sigma(r,\theta)\equiv 
\frac{\Sigma(r,\theta+\Omega_{\rm bin}t)}{\langle\Sigma\rangle(r)}-1
\eeq 
which selects the relative nonaxisymmetric component of the surface density
distribution. The fluctuation is a two-armed tightly wound spiral density
wave.  Since the wave is stationary in the rotating frame, its pattern
angular speed in the inertial frame equals $\Omega_{\rm bin}$.  At large
radii $r\gg 2a$, the wave obeys the Lin-Shu dispersion relation
\beq
\Omega^2-m^2(\Omega-\Omega_{\rm bin})^2+k^2c_{\rm s}^2=0
\eeq
with $m=2$, where $k$ is the radial wavenumber.

The wave appears to be excited in the region $a\lesssim r\lesssim 2a$ and
seems to refract at $r\sim 2.5a$ where the surface density profile exhibits a
jump.  The oscillating torque density in Figure \ref{fig:torque} directly
follows from the stationary spiral pattern in Figure \ref{fig:wave}.
Significant net torque is exerted only in the central, low-density hole and
near the disk edge ($r\lesssim 2a$) where the spiral wave is nearly radial;
the net contribution from larger radii is very small due to cancellation.

The total, time-averaged torque experienced by the disk is positive,
\beq
\label{eq:torque_measured}
T(\infty)\approx 6\times10^{-4}\ GMa\Sigma_0 .
\eeq  
The binary-disk torque can be compared to the viscous torque communicated in the disk
\bea
T_{\rm visc}&=&2\pi r^3 \nu \left\langle\Sigma \frac{\partial}{\partial r}\left(\frac{v_\phi}{r}\right)\right\rangle
\nonumber\\&\sim& 
- 3\pi\alpha\left(\frac{h}{r}\right)^2 GMr\Sigma .
\eea

The two torques are of comparable magnitude when $\Sigma/\Sigma_0\sim 0.64\
a/r$, which is roughly satisfied in the inner disk.  This is of course
expected if the disk has settled in a quasi-steady state characterized by a
balance of viscous and gravitational torques. But the balance is not exact,
because angular momentum is being transported by the (weak) radial accretion
and density waves in the disk.

\subsection{Fluid inside the Central Hole}

\begin{figure}
\epsscale{1.1}
\plotone{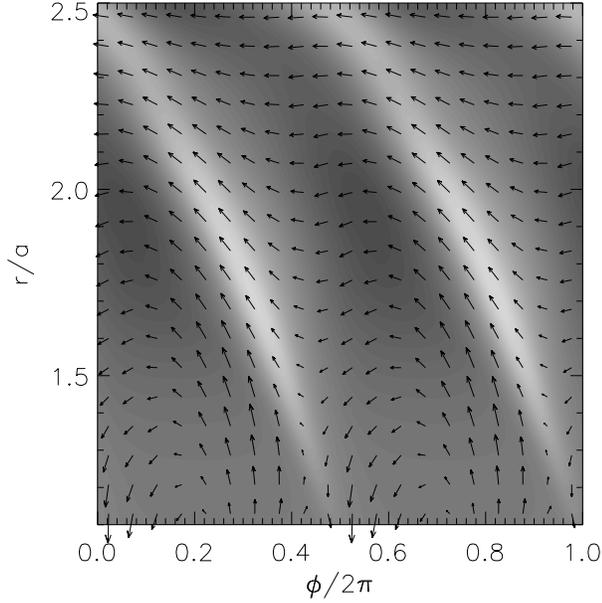}
\caption{Same as Figure \ref{fig:wave}, with the addition of vectors representing the time-averaged velocity of the fluid in the rotating frame in which the binary is at rest.  At $r\gg a$, the angular velocity approaches the negative angular velocity of the binary, $d\phi/dt\rightarrow -\Omega_{\rm bin}$.  The brightest regions are density maxima. The plot does not show the disk at radii $r_{\rm min}<r<1.1a$.\label{fig:wave_velocity}}
\end{figure}

Figure \ref{fig:wave_velocity} shows that the fluid flow inside the central
hole ($r\lesssim 2a$) is non-circular.  Of course, in this figure we only see
the high--pattern-speed $m=2$ component, whereas the low--pattern-speed $m=1$
component is stronger but not present after time averaging in the rotating
frame.  Examining the instantaneous flow within the central hole, we observe
pinball-like behavior; a fluid element located inside the hole receives a
gravitational kick near the pericenter and is ejected outward on an eccentric
orbit toward the inner edge of the disk.  It eventually collides with the
inner edge at $r\sim2a$.  The momentum of the ejected fluid is transferred to
the disk and a shock wave is driven into the disk.

In \S~\ref{sec:discussion}, we interpret the torque transfer by periodic
kicking of the eccentric fluid inside the central hole as a mean-motion
resonance.  In addition to the tightly-wound density wave observed in the
frame rotating with the binary, a strong, one-armed, long-wavelength density
wave is driven at the apocenter of the inner edge of the disk.  The latter
wave precesses with a low pattern speed and hence disappears when
time- veraging in the rotating frame.  The long-wavelength wave appears to be
associated with the propagation of the eccentric distortion in the disk.

\subsection{Accretion into the Central Hole}

The average mass accretion rate across the inner edge of the integration domain is 
\beq
\dot M(r_{\rm min})\approx 2.5\times10^{-4}(GMa)^{1/2}\Sigma_0.  
\eeq
This can be compared to the nominal accretion rate in a disk without
binary-induced torques
\bea
\dot M_{\rm free}&=&\left[\frac{d(r^2\Omega)}{dr}\right]^{-1}\frac{dT_{\rm visc}}{dr}\nonumber\\&\sim& 6\pi\alpha\left(\frac{h}{r}\right)^2 (GMr)^{1/2} \Sigma , 
\eea 
where we have assumed that in the disk without binary torque the surface density is a weak function of radius.  Then we have
\beq
\label{eq:mdot_ratio}
\frac{\dot M}{\dot M_{\rm free}}\sim 0.13 \left(\frac{r}{a}\right)^{-1/2}
\left(\frac{\Sigma}{\Sigma_0}\right)^{-1} .
\eeq

The accretion rate ratio in equation (\ref{eq:mdot_ratio}) should be evaluated at the outer edge of the zone where the binary forcing is significant. This is the transition radius where the disk surface density derivative changes sign, $d\langle\Sigma\rangle/dr\sim 0$; outside this radius, angular momentum transport in the disk is governed by nongravitational, viscous torques only. Figures \ref{fig:density} and \ref{fig:torque} suggest that the transition radius is at $r\sim3\ a$ where $\langle\Sigma\rangle\sim 0.425\ \Sigma_0$.  With this, $\dot M/\dot M_{\rm free}\sim 0.18<1$, 
which indicates 
that the accretion rate into the central hole in the circumbinary
disk is smaller than the accretion rate in a disk without binary torque with comparable density. 

The surface density profile of the disk exterior to the density maximum at $r\sim 3a$ is slightly steeper than the profile $\Sigma\propto r^{-1/2}$ expected in a constant-$\dot M_{\rm free}$ disk.  However the surface density profile is significantly shallower than expected in a non-accreting, constant-$T_{\rm visc}$ disk.  In the absence of accretion through the central hole, the disk behaves like a decretion disk \citep{Pringle:91}, with surface density profile $\Sigma\propto r^{-1}$.   This  indicates that the outer disk has not yet settled in the viscous quasi-steady state; the simulated disk is neither exactly an accretion nor a decretion disk.

\begin{figure}
\epsscale{1.1}
\plotone{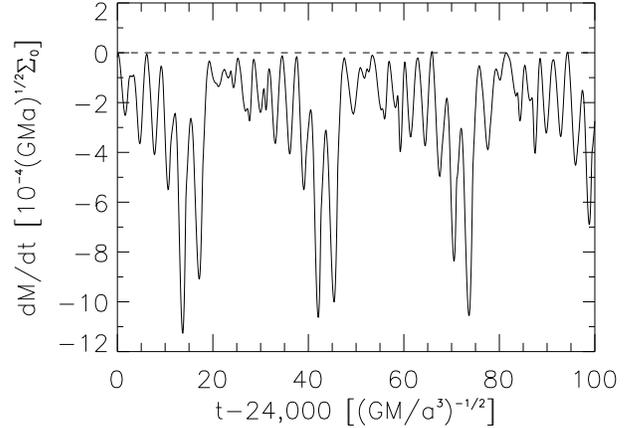}
\caption{Time dependence of the accretion rate $dM/dt=\dot M=r \int v_r \Sigma d\phi$ across the inner edge of the integration domain, $r=r_{\rm min}=a$, at $\sim 4,000$ binary revolutions after the beginning of the simulation.\label{fig:mdot}}
\end{figure}

Figure \ref{fig:mdot} shows the time dependence of the accretion rate.  This is the rate with which fluid flows across the inner boundary of the simulated domain; this flow should ultimately accrete onto either of the point masses or be otherwise removed (unbound) from the system.  The accretion rate is quasi-periodic and exhibits notable low-frequency variation and periodic outbursts. The peak accretion rate is a factor of $\sim 4$ larger than the mean value. We discuss the implications of the time-dependent accretion rate in \S~\ref{sec:black_holes}.

\begin{figure}
\epsscale{1.1}
\plotone{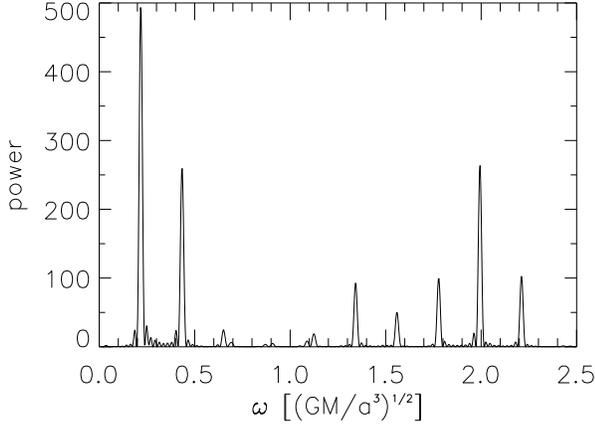}
\caption{The Lomb-Scargle periodogram of the rate of accretion across the inner boundary at $r=r_{\rm min}=a$.    The periodicity analysis was carried out over a $3$ times longer time interval than shown in Figure \ref{fig:mdot}.\label{fig:lomb}}
\end{figure}

Figure \ref{fig:lomb} shows the Lomb-Scargle periodogram of the accretion rate. The periodogram was calculated using the definitions in \citet{Scargle:82} and \citet{Horne:86}\footnote{http://astro.uni-tuebingen.de/software/idl/aitlib/timing/scargle.html}. The peak at $\sim 2\times\Omega_{\rm bin}$ corresponds to the forcing frequency of the quadrupole of the tidal potential. The accretion rate time series is quasi-periodic as it can be decomposed into 10 components with evenly spaced frequencies $\omega\sim \frac{2}{9} K\Omega_{\rm bin}$, where $K=1,2,\cdots,10$.  The peaks at $K=3,4$, and $5$ are not significant.  The most interesting is the fundamental frequency component at $\omega\sim\frac{2}{9} \Omega_{\rm bin}$; it is responsible for the slow periodic modulation of the signal seen in Figure \ref{fig:mdot}.

\subsection{Evolution of the Binary Separation}

The fluid accreting across the central hole transports angular momentum at the rate  
\beq
\dot L_{\rm acc}= 2\pi r^2\langle v_rv_\phi\Sigma\rangle .
\eeq
We measure the average $L_{\rm fluid}$ in the simulation and find
\beq
\dot L_{\rm acc}(r_{\rm min}) \approx 3.3\times10^{-4} GMa\Sigma_0 ,
\eeq
which is a half of the binary-disk torque (cf.~equation \ref{eq:torque_measured}).  Therefore the cross-hole accretion of (positive) angular momentum is not sufficient to offset the loss of angular momentum to the disk, and the binary loses angular momentum on average.  

If the binary remains circular,  evolution of the binary semimajor axis is governed by 
\beq
\dot L_{\rm acc} - T = \frac{(GM^3a)^{1/2}}{8}\left(\frac{\dot a}{a}+3\frac{\dot M}{M}\right) ,
\eeq 
where the minus sign in front of the torque on the left-hand side reflects $T$ being defined as the torque exerted by the binary on the disk.  The rate at which the binary separation decays can then be estimated to be
\bea
\label{eq:adot}
\frac{\dot a}{a} &\approx& -0.003\left(\frac{Ga}{M_{\rm bh}}\right)^{1/2} \Sigma_0\nonumber\\
&\approx& -7.3\times10^{-4} \left(\frac{M_{\rm edge}}{\onehalf M_{\rm bh}}\right) \frac{1}{t_{\rm bin}} ,
\eea
where $M_{\rm edge}\equiv \pi(2a)^2\Sigma_0$ and $M_{\rm edge}\ll M_{\rm bh}$ is assumed.  The inferred binary orbital decay time is comparable to the viscous time at the inner edge of the disk, multiplied by the mass of the disk at the inner edge and divided by the binary mass. If the binary does not remain circular, the estimate based on our simulation is not strictly accurate, but the overall scaling should still apply.  If the binary point masses are unequal, $\onehalf M_{\rm bh}$ in equation (\ref{eq:adot}) should be replaced by the reduced mass $M_1M_2/M_{\rm bin}$.

\section{Discussion}
\label{sec:discussion}

\subsection{Torque Density from Tidal Forcing at $m=2$ OLR}

The torque density profile in Figure \ref{fig:torque} resembles the theoretical torque density profile for a density wave linearly excited at an OLR by a binary's tidal potential in a circular, quasi--uniform-density disk \citep{MeyerVernet:87}.\footnote{We are grateful to Steve Lubow for providing this reference, and for suggesting an outline for the present subsection.}  To explore this similarity, we proceed to compare torques in the simulated disk with predictions of the linear response theory. In a circular, inviscid, non--self-gravitating, quasi--uniform-density Keplerian fluid disk that is forced by a rotating potential multipole $\Phi_m$ of order $m$, torque density in the vicinity of the corresponding OLR can be written in terms of the Airy function ${\rm Ai}(x)$ \citep{MeyerVernet:87}
\beq
\frac{dT}{dr} = \frac{T_m}{\lambda_m} {\rm Ai} \left(\frac{r_m-r}{\lambda_m}\right)
\eeq
where $r_m$ is the radius of the resonance that is the solution of 
the equation
$\Omega(r_m)=[m/(m+1)]\Omega_{\rm bin}$
and $T_m$ is the approximate total torque exerted at an OLR in a pressureless disk \citep{Goldreich:79}
\beq
T_m = m \pi^2 \left[\Sigma\left|r\frac{dD}{dr}\right|^{-1}
\left(r\frac{d\Phi_m}{dr}+\frac{2\Omega}{\Omega-\Omega_{\rm bin}}\Phi_m\right)^2\right]_{r=r_m} .
\eeq
Here, in a Keplerian potential, $D\equiv \Omega^2-m^2(\Omega-\Omega_{\rm bin})^2$, and $\lambda_m$ can be interpreted as the inverse radial wavenumber associated with the pressure wave excited at the resonance,
\beq
\lambda_m\equiv \left[r\left(\frac{c_{\rm s}^2}{3m r^2 \Omega\Omega_{\rm bin}}\right)^{1/3}\right]_{r=r_m} .
\eeq

The equal-mass binary analyzed here is special in that the potential component having $m=1$ that corotates with the binary is identically zero. Therefore, the outermost $m=1$ OLR at $r_1\approx 2^{2/3}\ a$ is absent.
The $m=2$ potential component associated with the outermost OLR at $r_2=(3/2)^{2/3}a$ with orbital resonance $\Omega_{\rm bin}:\Omega=3:2$, in the special case of an equal-mass binary, equals
\beq
\Phi_2 = \frac{3}{16} \left(\frac{r}{a}\right)^{-3} a^2\Omega_{\rm bin}^2 + {\cal O} \left[\left(\frac{r}{a}\right)^{-5}\right] .
\eeq
Note that although $r_2$ is the location of the resonance, the range of radii over which resonant forcing occurs is actually fairly broad, as is evident in Figure \ref{fig:torque}, and is comparable to the radius itself.
With this we obtain
\beq
\label{eq:torqueapprox}
\frac{dT}{dr} \approx \frac{49}{288}\pi^2\Sigma(r_2) \Omega_{\rm bin}^2 \frac{a^4}{\lambda_2}  {\rm Ai} \left(\frac{r_2-r}{\lambda_2}\right) ,
\eeq
where $\lambda_2=2^{-2/3} (h/r)^{2/3} a$.  

The approximate torque density derived from linear theory in equation (\ref{eq:torqueapprox}) is only a crude approximation, because the simulated disk violates multiple premises on the which the derivation is based:

1. The strength of the resonant torque associated with the $m=2$ OLR is nonlinear, $T_2/[a^4\Omega_{\rm bin}^2\Sigma(r_2)]\gtrsim 1$.  In particular, nonlinear coupling with the eccentric $m=1$ mode may modify the torque density.  At the radius of the resonance $r_2$, the disk itself is intermittent, rather than steady.

2. The derivation assumes that the forcing potential varies very slowly with radius, $d\ln\Phi_2/dr\ll 1/\lambda_2$, but in the simulation, the forcing potential and the resonantly-excited wave vary on comparable length scales.

3. The derivation assumes a quasi--uniform-density disk, $d\ln\langle\Sigma\rangle /dr\ll 1/\lambda_2$, but in the simulation, the azimuthally-averaged density  and the pressure vary more rapidly than the resonantly-excited wave. Since the resonance has a finite width, the resonant torque will depend on the density over a range of radii.

4. Similarly, the linear derivation assumes an isothermal disk, whereas our disk is locally-isothermal and the sound speed varies as $c_{\rm s}\propto r^{-1/2}$.

5. The derivation assumes a Keplerian potential, whereas the potential of the binary is stronger than Keplerian at small radii (e.g., see equation [\ref{eq:omega}]).

Each of these aspects can be, with some effort, incorporated in an analytic derivation of the torque density. Instead, here we simply attempt to evaluate the torque density estimate according to equation (\ref{eq:torqueapprox}) after substituting the disk surface density (equation [\ref{eq:sigmasmall}]) evaluated at the resonant radius $r_2$. The surface density equals $\Sigma(r_2)\approx 6\times10^{-4}\ \Sigma_0$, and with this, the estimated torque density reaches maximum 
\beq
\label{eq:torquepeak}
\left(\frac{dT}{dr}\right)_{\rm max}\approx 5.5\times10^{-4}\ a^3 \Omega_{\rm bin}^2 \Sigma_0
\eeq
at radius $r\approx 1.45\ a$.  The peak torque density in equation (\ref{eq:torquepeak}) is a factor of $\sim2.5$ higher than the peak torque density in the simulation, which occurs at radius $r\approx 1.75\ a$ (cf.\ Figure \ref{fig:torque}).  At larger radii, the theoretical torque density oscillates more rapidly than the simulated, that is, the wave excited at the resonance has a shorter wavelength (e.g., the fifth positive torque peak occurs at $r\approx 2.87\ a$ in linear theory, and at $r\approx 3.55\ a$ in the simulation).    This is not surprising, given that linear theory can severely underestimate the radial wavelength of the density wave excited at an OLR when the forcing is nonlinear \citep{Yuan:94}. We can conclude that the torque density measured in the simulation is consistent with having arisen from a forcing by the binary's tidal potential at the $m=2$ OLR.  This implies that the same resonance can be responsible for disk truncation and maintenance of torque balance near the inner edge of the disk.

\subsection{Qualitative Interpretation of Disk Eccentricity Excitation}

We here provide a qualitative interpretation of the excitation and maintenance of disk eccentricity.  Consider a disk fluid element orbiting the binary on a quasi-elliptical orbit with semimajor axis $\tilde a$ and eccentricity $\tilde e$.  The fluid element makes a pericenter passage at radius $r_{\rm p}\sim (1-\tilde e) \tilde a$.  Its angular frequency at the pericenter is approximately given by
\beq
\Omega_{\rm p}\sim \Omega_{\rm kep}(\tilde a) \left[1+\frac{3}{16} \frac{a^2}{(1-\tilde e)^2\tilde a^2}\right]\frac{(1+\tilde e)^{1/2}}{(1-\tilde e)^{3/2}}.  
\eeq
When this frequency is close to the orbital frequency of the binary
$\Omega_{\rm bin}$, the fluid element experiences a gravitational pull as it passes the pericenter point.  Assuming that the eccentricity at the pericenter  is $\tilde e\sim 0.2$ (see Fig.~\ref{fig:eccentricity}), the resonance condition 
\beq
\label{eq:resonance}
\Omega_{\rm p}(\tilde a,\tilde e)=\Omega_{\rm bin}
\eeq 
can be solved to obtain the numerical solution $\tilde a\approx 1.45\ a$. 

The fluid that arrives with a slightly larger $\tilde a$ and slightly smaller
$\tilde e$ progresses in azimuth at a slightly lower rate than the binary
point mass during pericenter passage.  In this case, the binary  imparts a
gravitational ``kick'' to the fluid element.  Transfer of momentum from the
binary to the disk is evident in the positive torque density inside the
central hole occupied by orbits close to the resonant condition in equation
(\ref{eq:resonance}), as seen in Figure \ref{fig:torque}.  The density wave
(Figure \ref{fig:wave}) is clearly excited inside the hole.

After departing the pericenter and proceeding to larger radii, the orbital frequency of the fluid decreases.  In a time set by the relation
\beq
\int_0^t [\Omega_{\rm bin}-\Omega(r(t'))] dt' \sim \frac{\pi}{2}
\eeq 
the kicked fluid  lags by a quarter of a phase behind the binary, and the sign of the torque  changes, because now the other point mass is located closer to and just behind the kicked fluid.  Therefore, a negative torque density is expected at larger radii, as seen in Figure \ref{fig:torque}. 

The point mass approximation employed thus far breaks at the apocenter.  The
simulation shows that the kicked fluid impacts the inner edge of the disk
near the apocenter supersonically which drives a shock wave into the disk.
The kicked fluid imparts momentum to the disk which is then transported into the disk interior by a density wave.  Therefore, the low-density fluid located in the central hole that experiences periodic kicks mediates the binary-disk torque coupling.  

After the impact at the apocenter, the kicked fluid loses orbital phase
coherence, and plunges back toward the pericenter as an eccentric stream.
Consecutive kicks do not add up coherently; the fluid does not retain the
memory of its previous pericenter passages, which distinguishes it from the
collisionless eccentric mean motion resonance studied by \citet{Pan:04}.
This puts into question the assumption at the heart of the standard treatment
of binary-disk resonances that disturbances are linear and that consecutive
disk-planet passages can interfere coherently.  Dispersal of the wave excited
by each kick implies that the binary must resonate with the angular frequency at
the pericenter alone, but not with any of the lower frequency components
associated with the eccentric disk.

The application of a positive torque near the pericenter and a negative
torque near the apocenter acts to fortify the eccentricity of the inner disk.
We speculate that this mechanism, in a linear form, is responsible for the
transition from an elliptical to an eccentric disk.  
Indeed, in an equal-mass binary,
disk ellipticity can be linearly excited at the 2:1 ``eccentric'' OLR 
($r\approx 1.59 a$) by the mechanism derived in
\citet{Lubow:91} and \citet{Dangelo:06}. 
 
Finally, we mention an alternative historical approach to disk truncation in
circumbinary disks which is based on the behavior of the restricted
three-body problem.  The growth of orbital eccentricity of a collisionless
third body orbiting a similar-mass binary occurs in the vicinity of
parametric resonances \citep{Henon:65,Szebehely:67,Rudak:81}. If the orbit of
the third body is interpreted as representing a streamline in the fluid
circumbinary disk, the catastrophic self-intersection of resonantly-excited,
eccentric orbits leads to fluid removal and disk truncation
\citep{Rudak:81,Erwin:99,Pichardo:05}.  We are not aware of an explicit
demonstration that an equivalent of the three-body parametric resonances can
truncate a fluid circumbinary disk.

\section{Circumbinary Disks in Massive  Black Holes}
\label{sec:black_holes}

A massive  binary black hole forms on a dynamical timescale at the conclusion of a galaxy merger, but can then  persist at parsec or subparsec separations over a cosmological time \citep{Begelman:80}.  Initially, the binary  decays by transferring orbital energy and angular momentum to the stellar environment.  However when the separation between the black holes becomes (\citealt{Peters:64}; a circular binary is assumed)
\bea
\label{eq:a_gr}
a_{\rm gr}(t) &\sim& 2500\ r_{\rm g}
\left(\frac{t}{10^6\textrm{ yr}}\right)^{1/4}
\left(\frac{M}{10^6 M_\odot}\right)^{-3/4}
\nonumber\\
&\times&
\left(\frac{M_1}{5\times10^5 M_\odot}\right)^{1/4}
\left(\frac{M_2}{5\times10^5 M_\odot}\right)^{1/4} ,
\eea
where $t$ denotes a time interval and $r_{\rm g}\equiv GM/c^2$ with $M\equiv M_1+M_2$, the emission of gravitational radiation drives the black holes to coalesce in time $t$.  Note that the dependence on $t$ and the total black hole mass is weak.  In the gravitational-radiation--dominated phase, the binary will be very close to circular.

Consider a binary composed of low-mass massive black holes ($M\lesssim 10^7M_\odot$).  Equation (\ref{eq:a_gr}) then implies that  the binary spends time $t\gtrsim 10^6 \textrm{ yr}$ at separations $a\lesssim a_{\rm gr} \sim (10^3-10^4) r_{\rm g}$.  At such separations, the binary is sufficiently compact for it to be possible that a hot and ionized ($T\gtrsim10^4\textrm{ K}$) circumbinary accretion disk can encompass it.  This type of accretion disk forms when interstellar gas accretes near a pre-existing binary.  The size of the disk is limited by self-gravity at the outer edge \citep[and references therein]{Kolykhalov:80,Goodman:03}.   

Since the circumbinary disk contains a central hole and the accretion across the hole and onto the individual black holes is reduced, the disk luminosity must be calculated self-consistently in any invocation of the Eddington limit.  The standard ``Eddington'' accretion rate, calculated assuming a luminosity $L\sim \epsilon \dot M_{\rm free} c^2$, where $\epsilon$ is a dimensionless efficiency factor, is of no relevance to the circumbinary disk. At binary separations $a\lesssim 100r_{\rm g}$, the binary decay timescale becomes shorter than the viscous response time of the disk, and the two components decouple \citep{Milosavljevic:05}.

Besides likely containing an unequal mass binary ($M_1\neq M_2$), the astrophysical hot circumbinary disk will differ in two significant aspects from the model disk in our simulation.  The midplane temperature in the real disk at radii $r\sim (100-1000)r_{\rm g}$ will be in the range $10^4-10^6\textrm{ K}$ (e.g., \citealt{Hubeny:01}), implying  a much cooler and thinner disk ($h/r\sim 0.001-0.01$) than in our simulation ($h/r\sim 0.1$).  Furthermore, the locally isothermal equation of state assumed in our simulation will almost certainly be a poor approximation in the real disk; the inner edge of the disk will be heated by a combination of viscous torques and dissipation of waves driven by the binary.  Nonetheless, the basic dynamical characteristics identified in the simulation should be present in the real disk; it is thus important to examine their implications for the detectability of compact massive black hole binaries embedded in circumbinary disks.\footnote{An important caveat is our assumption that the disk is corotating with the binary. The dynamics of a counter-rotating disk will be very different and must be studied separately.}

The principal challenge to astronomical detection of an active galactic nucleus (AGN) containing a binary massive black hole is the confusion with normal AGNs that contain a single black hole.  Our simulation  suggests multiple forms of variability that can serve as indicators of a binary black hole in an AGN. The variability will be associated with: (1) the propagation and dissipation of density and shock waves in the bulk of the circumbinary disk; (2) supersonic gas dynamics in the central hole in the disk; (3) transfer of material across the hole onto the black holes; and (4) ultimate accretion of the material that has crossed the hole and has become captured by one of the black holes onto the black hole, which may also proceed via a thin disk.    Detailed modeling of the emission from these structures must be done to identify the most promising indicators of binary massive black hole AGNs (e.g., \citealt{Bogdanovic:07b}). 

The amplitude of the variability produced by the strong supersonic shocks within the central hole can be estimated to be $L_{\rm hole}\sim M_{\rm hole} v^2 \Omega_{\rm bin}$, where $M_{\rm hole}\sim \alpha (h/r)^2 M_{\rm edge}$ is the mass of the gas inside the hole, $v\sim (GM_{\rm bh}/a)^{1/2}$ is the orbital velocity, and as before $M_{\rm edge}$ is the mass of the disk at the inner edge (we ignore dimensionless factors of order unity).  This can be compared to the accretion luminosity of the disk $L_{\rm disk}\sim G M_{\rm bh} \dot M_{\rm visc}/a \sim \alpha (h/r)^2 (G M_{\rm bh} /a)M_{\rm edge}\Omega$.  Evidently $L_{\rm hole}\sim L_{\rm disk}$, but the respective spectra will be considerably different; the emission inside the central hole will be harder than that of the disk.

\citet{Artymowicz:96} suggest that mass transfer from a circumbinary disk onto an eccentric binary will be punctuated by periodic outbursts.  We, however, find that the mass transfer (see Fig.~\ref{fig:mdot}) is quasi-periodic and highly variable even if the binary is circular; this is the case because the disk itself becomes eccentric.  Temporal modulation of the cross-gap accretion in an  eccentric protoplanetary disk is simultaneously being reported by \citet{Dangelo:06}, who simulate binaries with very unequal-mass components, $M_2/M_1\sim 0.001-0.003$.  In a similar-mass binary, the principal period is half of the binary orbital period
\bea
\frac{t_{\rm bin}}{2}&\sim& 23\textrm{ days}\ \left(\frac{t}{10^6\textrm{ yr}}\right)^{3/8}
\left(\frac{M}{10^6 M_\odot}\right)^{-1/8}
\nonumber\\
&\times&
\left(\frac{M_1}{5\times10^5 M_\odot}\right)^{3/8}
\left(\frac{M_2}{5\times10^5 M_\odot}\right)^{3/8} ,
\eea
but longer periodicities (e.g., $\sim 4t_{\rm bin}$) marked by outbursts of  accretion across the hole can also be expected.  The variability will be present on optical to X-ray wavelengths and can be detected by all-sky monitoring surveys such as LSST.  

Accretion modulated by an eccentric binary companion (e.g., \citealt{Sillanpaa:88,Lehto:96,Valtaoja:00}) has been invoked as an explanation of the $12\textrm{ yr}$ optical periodicity in the BL-Lacertae active galaxy OJ 287 (e.g.,~\citealt{Sillanpaa:96a,Sillanpaa:96b,Pursimo:00,Valtonen:06}). The optical light curve of OJ 287 is punctuated by tandem flares separated by $\sim 11\textrm{ yr}$  of relative quiescence.  The binary black hole models of OJ 287 involve a single or double crossing of one black hole's accretion disk by the other black hole; the semimajor axis of the black hole binary is assumed to be larger than the accretion disk.  

The quasi-periodic accretion across the central hole in our simulation (Fig.~\ref{fig:mdot}) tentatively suggests an alternative explanation for the structure of the light curve of OJ 287.  Evidently, a modulation of the accretion rate resulting in twin outbursts can take place in a circular binary surrounded by a coplanar disk.  However a model of OJ 287 based on a circumbinary disk must explain the puzzling absence of a radio counterpart in  the  first of the two optical flares \citep{Valtaoja:00}.

Finally we remark that if the disk is sufficiently massive while of course remaining gravitationally stable and geometrically thin, the torque imparted by the disk on the binary can compete with stellar dynamical processes at extracting energy and angular momentum from the binary and accelerating coalescence (e.g.,~\citealt[and references therein]{Ivanov:99,Gould:00,Merritt:05}).  At present we are unable to estimate the expected mass of such a disk from first principles.  \citet{Armitage:05} have shown that  disk torques can excite eccentricity in a binary that has entered the gravitational-radiation driven inspiral.  We speculate that the disk eccentricity that develops on a viscous timescale in our simulation may contribute to the excitation of the binary's eccentricity; \citet{Papaloizou:01} have shown that the two eccentricities are coupled.  Recently, \citet{Bogdanovic:07a} pointed out that cross-gap accretion will torque the angular momenta of the two black holes to align them with that of the disk, which may restrict the details of the ultimate gravitational-radiation mediated coalescence to the regime not propitious to strong gravitational recoil.  This suggests that the consequences of geodetic precession that is present if the black holes are rapidly rotating should be explored in future extensions of our study.

\section{Conclusions}
\label{sec:conclusions}

We carried out a grid-based, two-dimensional hydrodynamic simulation of a thin, locally-isothermal, corotating, non--self-gravitating, viscous accretion disk around a Newtonian equal-mass binary on a fixed circular orbit.  The disk is evolved over multiple viscous times of the inner disk.  We study the quasi--steady-state structure of the circumbinary fluid flow and the mechanics of the binary-disk torque coupling.

We here summarize the main conclusions of this work.

1. Fluid density within twice the binary's semimajor axis is significantly
   reduced.  The point masses orbit within the central low-density hole in
   the disk.  The disk first becomes elliptical and then eccentric.  The
   eccentricity of the disk is strong and precesses slowly.  Radial
   fluid motion within the central hole is supersonic and
   time-dependent. Strong shocks are driven into the low-density fluid by the
   tidal field.

2. The binary exerts torque on the disk by imparting gravitational kicks to
   the eccentric low-density fluid inside the central hole.  Angular momentum
   is transferred to the disk as the kicked fluid impacts the edge of the
   hole.  A two-armed spiral density wave that is stationary in the
   rotating frame in which the binary is at rest is driven into the disk.
   The structure of the two-armed density wave and that of the binary 
   torque distribution in the disk suggest that the binary forces the disk 
   at the $m=2$ outer Lindblad resonance, where the orbital 
   period of the disk is $3/2$ times the binary's orbital period.

3. The accretion into the hole toward the two point masses is reduced
   relative to the rate expected for a disk with the same surface density and
   a torque-free inner edge.  However, significant mass transfer from the
   disk onto the black holes does take place.  The mass transfer rate is
   quasi-periodic and punctuated by outbursts.  The accreting fluid carries
   angular momentum, but the amount of angular momentum accreted onto the
   binary is insufficient to offset the loss of angular momentum to the
   gravitational torque.  If the binary remains circular, its separation
   decays on a viscous time of the disk reduced by the ratio of the mass at
   the inner edge of the disk to the binary mass.

4. The quasi-periodic accretion across the hole, the shocks in the
   low-density fluid inside, and the disk eccentricity are sources of
   variability that can be used to distinguish an AGN containing a binary
   massive black hole from a regular AGN containing a single black hole.
   Detailed modeling of the emission from these structures is required to
   identify the most promising indicators of binary massive black hole AGNs.

\acknowledgements

We would like to thank Steve Lubow for 
comments that have helped improve our 
understanding of the mechanics of binary-disk
 coupling in the simulation.  We would also like to thank
 Tamara Bogdanovi\'c for comments and 
Phil Armitage, Mitch Begelman, Roger Blandford, Paolo Coppi, Peter Goldreich,
Jim Gunn, Margaret Pan, 
Sterl Phinney, Roman Rafikov, and Mateusz Ruszkowski for helpful conversations.  
We would also like to thank Michael Turk for his help during an
early phase of the project.  A.~I.~M.\ acknowledges support from a Keck
fellowship at the Institute for Advanced Study.   M.~M.\  acknowledges support by NASA through the Hubble
Fellowship grant HST-HF-01188.01-A awarded by the Space Telescope Science
Institute, which is operated by the Association of Universities for Research
in Astronomy, Inc., for NASA under contract NAS5-26555. 
The software used in
this work was in part developed by the DOE-supported ASC/Alliance Center
for Astrophysical Thermonuclear Flashes at the University of Chicago. 
The simulations presented in this work were performed with
the Scheides cluster at the Institute for Advanced Study.
M.~M.\ thanks the Kavli
Institute for Theoretical Physics for hospitality during the completion of
this work. This
research was supported in part by the National Science Foundation under grant
PHY99-07949.

\end{document}